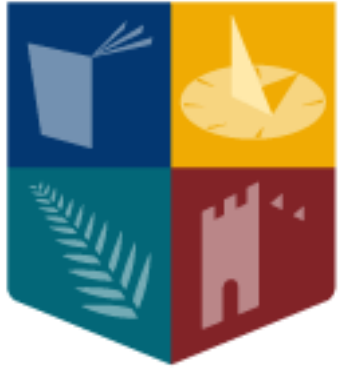

# Using an innovative assessment approach on a real-world group based software project

Susan Bergin and Aidan Mooney

Department of Computer Science, Maynooth University,
Maynooth, Co. Kildare, Ireland
Date: July 2016

susan.bergin@nuim.ie
aidan.mooney@nuim.ie

# Table of Contents

# Abstract

Currently, there is a lack of practical, real-world projects on Computer Science (CS) courses at Maynooth University. Generally CS undergraduate modules are composed of 24 hours of lectures and 24 hours of labs where students learn theoretical concepts in the lectures and apply their understanding to practical lab-based exercises. The problem with this approach is that students do not gain any awareness of, or learn how to solve tasks that they are likely to encounter in a real-world industrial setting; nor do they gain experience of working as part of a team even though most software development positions involve team-based work.

This paper reports on a web-based development module that incorporated a real-world group based project was re-designed and delivered. The module went well; however, assessing the work fairly was found to be difficult, especially where team members contributed at considerably varying levels was a challenge. Of particular concern was that some hard-working students were penalised by other students' poor work and lazy students were rewarded because of more hard-working students' work.

This action research project will attempt to re-address how to assess this group-based work with a cohort of students. The goal of the research is to implement an innovative assessment structure, using peer-, self-, and co-assessment, for a group based real-world project, that is deemed fair and reasonable and provided a good learning environment.

# 1. Educational Situation & Literature Review

## *1.1 Background*

CS230 Web Information Processing is an undergraduate module on WWW technologies in our department. The advantage of this module is that it is very practical and opportunities exist to give student's exposure to projects they might work on in an industrial setting. This is important as currently, there is a lack of practical, real-world projects on our CS degree programmes. Typically, CS modules are composed of 24 hours of lectures and 24 hours of labs where students learn theoretical concepts in the lectures and apply their understanding to practical lab-based exercises. The problem with this approach is that students do not gain any awareness of, or learn how to solve tasks that they are likely to encounter in a real-world industrial setting nor do they gain experience of working as part of a team even though most software development positions involve team-based work. Furthermore, from an employer's perspective there is a growing concern on the quality of recent graduates and moreover on the necessity to re-train graduates as they have gained so few employment ready skills.

To address this issue in part the CS230 module was re-designed to incorporate a real-world web-based group project. Students participating on this module typically are studying on five different programmes: second year students on our dedicated Computer Science and Software Engineering degree (CSSE), second year BA / BSc Multimedia students, Higher Diploma in Information Technology students and visiting international students. This blend of technical, creative, mature and international students provides welcome diversity to the module as designing web pages requires artistic flair, technical ability and awareness of different user preferences and types.

## *1.2 Review of Related Literature*

Broadly speaking there are two main goals of assessment: (1) to provide certification of academic performance and (2) to improve the quality of student learning (OECD, 2013; Goode, 2010; Sluijsmans *et al.,* 1999). The first goal is largely achieved through summative assessment (identifying and grading what a student has learnt) whilst the latter is typically achieved by formative assessment (providing feedback of good quality information so that the learner can benefit from it) (Boud, 1998). Evaluation and assessment should align with the principles embedded in educational goals (OECD, 2013).

Current graduates are often expected to work in teams, make decisions and handle responsibilities in dynamic work environments with often non-routine abstract tasks. Graduates need to be able to analyse information, problem solve, communicate and reflect on their own role and performance.  Our role in higher education must be to help prepare students for taking up such positions and thus provide opportunities to facilitate the development of higher order life learning skills (Van den Bergha *et al.,* 2006, Boud, 1990). Alternative learning methods and environments such as group based work and project-based learning provide for such opportunities. Such approaches fall under the category of constructivist learning where learning is actively constructed by the learner and is self-regulated, goal-driven, contextual and often collaborative.

O'Farrell (2002) suggests that when designing and carrying out assessment it is important that the teachers and the students are clear as to what is expected of the student, along with the marks to be awarded. Students sometimes feel that the assessment criteria are the property of the teacher. However, there is no need for secrecy as being upfront about the assessment will direct students to what is expected of them and consequently will lead to much deeper learning.

According to Tucker *et al.* (2007) the incorporation of group based work has increased in higher education. This is driven by the commonly accepted arguments that peer learning can improve the overall quality of student learning and group work can help develop specific generic skills sought by employers such as critical enquiry, reflection ability and communication skills that are not as easily developed through more traditional approaches (Tucker *et al.,* 2007; Boud *et al.,* 1999).

Group assessment occurs when individuals work collaboratively to produce a piece of work. The advantage of group work for the assessor is often that the burden of marking many individual pieces of work is significantly reduced (DIT, 2008). A significant concern reported by students (Tucker *et al*., 2007) is that group work is fairly assessed and that individual contributions are justly rewarded. Boud *et al.* (1999) maintains that assessment is the single most powerful influence on learning in formal courses and in a collaborative learning environment and students must perceive the assessment methods to be credible and transparent. Schemes in which there is an explicit mix of individual and group assessment for common tasks may help to alleviate this problem. Tucker *et al.* (2007) contends that methods such as self-assessment and peer-assessment are valid and reliable alternatives to teacher-only assessment of individual contributions to group work.

**Self-assessment** is concerned with learners making judgments about their own learning and achievements (Falchikov and Boyd, 1989). By doing this students' take responsibility and become more actively involved in their own learning process thereby developing their own skills as reflective practitioners capable of lifelong learning.

Falchikov (2000) describes **peer-assessment** as being concerned with the process in which groups or individuals rate the performance of their peers on instruments designed by third parties or by the students themselves. The advantages of this approach include increasing student responsibility and involvement in the process and providing students with insight on the criteria determining the quality of their own work as well.

In **co-assessment** students and the staff collaborate in the assessment process. Both parties work together to define a mutually agreed assessment of the student's knowledge. This approach enables students to become active players in the assessment process whilst allowing staff to maintain a certain degree of control over the final assessment (Kilic, 2016; Hall, 1995). In a project-based learning study carried out by Van den Bergha *et al.* (2006) it was found that students believed that co-assessment allowed them to have a certain collaborative involvement in the assessment process; that it provided a happy medium between traditional and alternative modes of assessment and appreciated that it gave them the opportunity to defend or justify themselves.

Boud (2000; 1997; 1990) argues that an important aspect of assessment is the ability of students to monitor their performance and make assessments of what they need to do. Learning logs or reflective journals can be used by students to reflect on their learning process (Park, 2003; Francis, 1995). Students typically use these logs to summarise how they felt about their learning experience and draw conclusions on the process. In a study by Van den Bergha *et al.* (2006), it was found that students perceived the use of reflective journals as one of the most effective assessment tools on a group based project. They believed that it provided the instructor with clearer insight into the internal group and gave students the opportunity to give feedback to the instructor and to justify their actions. The students considered the reflective journal to be mainly a formative instrument, but were not opposed to it being graded.

Numerous studies highlight the benefits of self-, peer- and co-assessment methods. Siow (2015) states that students feel that incorporating self-assessment in their assignment makes them independent learners, think and learn more, become critical thinkers, work in a structured way, and become analytical. Tucker *et al.* (2007) described benefits such as promoting effective teamwork, developing professional skills in self-reflection on behaviour, developing graduate attributes for working in multidisciplinary teams and lifelong learning and shifting the student's role from passive receiver to active participant. Dochy *et al.* (1999) identified numerous positive effects of these methods in improving the quality of learning of students: increased student confidence in the ability to perform, increased awareness of the quality of the student's own work, increased student reflections on their own behaviour and/or performance, increased student performance on assessments, increased quality of the learning output and Increased student satisfaction. In addition, Sluijsmans *et al.* (1999) identifies the strengths of self-, peer-, and co-assessment methods as: development of student ownership of their own learning, motivating students and facilitating active involvement, encouraging students to become more autonomous learners, development of transferable skills and showing students that their experiences are valued and their judgments are respected.

Boud (1998) recommends that well-designed assessment tasks are authentic and set in a realistic context; are worthwhile learning activities in their own right; permit a holistic rather than a fragmented approach, that is, they engage students in the whole of a process; are not repetitive for either student or assessor; prompt student self-assessment; are sufficiently flexible for students to tailor them to their own needs and interests and are not likely to be interpreted by students in a way fundamentally different to those of the designer. A method to help achieve this is constructive alignment.

Constructive alignment, put forward first by Biggs (1996), is the aligning of all the components in a learning system such as the learning outcomes, the teaching methods and the assessment tasks so that the learning activities should lead to the desired learning outcomes (Hurley Lawrence, 2009; Biggs 2003). The role of the instructor is to create a learning environment that supports the learning and assessment activities appropriate to achieving the desired learning outcomes. Typically, such an approach has four main steps: define the intended learning outcomes; choose teaching and learning activities that are likely to lead to the desired learning outcomes; assess students' actual learning outcomes to see how well they match what was intended and arrive at a final grade.

Building upon the educational situation described in this section and the literature review on assessment, the goal of this research study is consequently to implement an ***innovative assessment structure, using peer-, self-, and co-assessment,*** for a ***group based real-world project,*** that was deemed ***fair and reasonable*** and provided a good ***learning environment***.

# 2 Methodology
To provide a sound theoretical grounding this project implemented an Action Research approach.

## *2.1 Action Research*
According to Carr (1986) Action Research is a form of self-reflective enquiry undertaken by participants (teachers, students or principles, for example) in social (including educational) situations in order to improve the rationality and justice of: (a) their own social or educational practices, (b) their understanding of these practices, and (c) the situations (and institutions) in which these practices are carried out.

Kurt Lewin is generally considered the 'father' of Action Research. Lewin first coined the term 'Action Research' in his 1946 paper "Action Research and

Minority Problems", characterising Action Research as a comparative research on the conditions and effects of various forms of social action and research leading to social action, using a process of a spiral of steps, each of which is composed of a circle of planning, action, and fact-finding about the result of the action (Lewin, 1946).

Action Research is any systematic inquiry conducted by teachers and other educational professionals in teaching/learning environments to gather information and reflect upon how their school operates, how they teach, or how well their students learn. Information is gathered with goals including effecting positive changes in the classroom and school environments, and improving student outcomes (Mills, 2003). Action Research encourages teachers to be participant researchers to gather information to share with the educational team. This information allows immediate analysis of instructional and behaviour management issues and is used to develop the next step(s) Teachers draw from research design tools that will describe what they are seeing in order to analyse and develop solutions and thereby improve their practice.

Stephen Kemmis designed a scheme to model the cyclical nature of a typical action research process - as shown in Figure 1. Each cycle has four steps: plan, act, observe, and reflect.

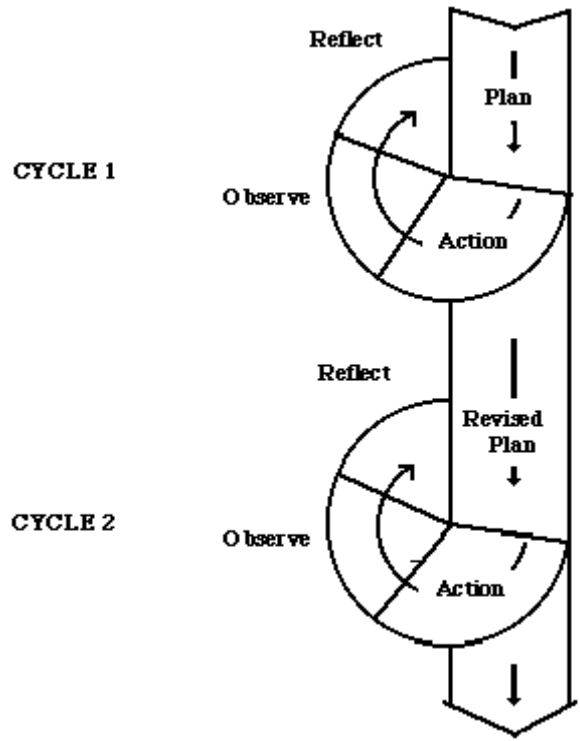

**Figure 1: Simple Action Research Model (taken from O'Brien, 1998)**

Similarly, Kurt Lewin (1948) described the process in terms of planning, fast finding, and execution. Planning starts usually with something like a general idea. The first step is to examine the idea carefully in the light of the means available. Frequently more fact finding about the situation is required. If this first period of planning is successful, two items emerge: namely, an 'overall plan' of how to reach the objective and secondly, a decision in regard to the first step of action. Usually this planning has also somewhat modified the original idea.

The next period is devoted to executing the first step of the overall plan and this second step is followed by certain fact-findings. This reconnaissance or fact finding has four functions. First it should evaluate the action. It shows whether what has been achieved is above or below expectation. Secondly, it gives the planners a chance to learn, that is, to gather new general insight.

Thirdly, this fact-finding should serve as a basis for correctly planning the next step. Finally, it serves as a basis for modifying the 'overall plan'.

The next step again is composed of a circle of planning, executing, and reconnaissance for the purpose of evaluating the results of the second step, for preparing the rational basis for planning the third step, and for perhaps modifying again the overall plan.

There is general agreement on the basic steps of an action research study:

    (1): identify an area of focus or concern,

    (2): collect data for documentation,

    (3): analyse and interpret data, and

    (4): share the information with others and develop an action plan (Arhar *et al.*, 2001; Schoen and Nolen, 2004; Stringer, 2004; Mills, 2003).

As noted by Mills (2003), action research uses elements of quantitative (e.g. comparison of standard scores) and qualitative research methods. However, the literature emphasises the data collection tools of qualitative research. These include use of observation, interviews, questionnaires, checklists, rating scales, focus groups, records, videotape, audiotape, and photographs (Arhar et al., 2001; Mills, 2003; Stringer, 2004).

### *2.2 Module Description*

The continuous assessment component of the CS230 module (worth 50%) required students to develop a website to promote the study of Computer Science at third level and specifically at Maynooth University. At the first lecture the students were asked to rate their current experience of the various technologies that would be used (XHTML, JavaScript, CSS, PhP etc). From this nine groups of five to six students were created within the class. As far as possible each group had a blend of previous experience of the technologies and had at least two CSSE students, two multimedia students and one HDipIT / international student. It was felt that this mix of students was fair and should lead to interesting discussion, design and final results.

Each group was asked to work together to complete a number of tasks over the duration of the module, including:

- Task 1: Promote the study of Computer Science at Third Level.
- Task 2: Promote Maynooth University and the CS Department.
- Task 3: Develop an interactive tool to allow prospective students to learn a CS topic.
- Task 4: Do something unique related to CS or to promote CS, Maynooth University or the Department.
- Complete group and student learning logs for each task.

In addition, students were awarded marks for formal and informal oral presentations / discussions during the project. An assessment structure that incorporated self-assessment, through reflective learning logs, peer assessment through peer-marking of presentations and co-assessment by developing marking schemes for allocating marks for the various parts of the project with the students was constructed. The format of the assessment structure is outlined next.

## *2.3 Implementation of co-assessment*

During the project the students and the lecturer worked together to co-design three marking schemes in the following order:

- Marking scheme for presentation
- Marking scheme for Task 1 and Task 2
- Marking scheme for Task 3 and Task 4

For the first marking scheme each group was given an identical incomplete marking scheme template and asked to modify and complete the scheme during a lab session. Each group had a mentor (postgraduate student) who assisted them but did not influence the criteria or marks they came up with. Upon completion the nine schemes were collated into a final marking scheme. To this end, it was identified where groups used different wording but appeared to be describing the same criterion and this determined the number of occurrences for each criterion. Marks were allocated proportionately to the number of occurrences. A small number of criteria which were believed to be unsuitable were not included in the final scheme. The students received (1) a copy of each group's marking scheme, (2) a spreadsheet detailing each criteria and occurrences, and (3) the final marking scheme so they could confirm that the final scheme was in fact reliably based on their work.

For the second marking scheme each group prepared a scheme from scratch. Again a spreadsheet was created detailing each criteria and the number of occurrences. During one of the lectures the students were given a copy of this spreadsheet and it was displayed to the class. Discussions between the lecturer and students occurred where each criterion was discussed and collectively an appropriate mark for each was decided upon. Considerable discussion arose in the classroom with groups being asked to explain why they wanted the criterion included with other groups arguing against its inclusion at times. Where a clear decision couldn't be reached a vote was taken and the majority decision held. Only once did the lecturer feel it necessary to overrule a decision and that was because it was felt that the students were allocating too few marks to a piece of work that had taken them a considerable length of time. The atmosphere was very relaxed and jovial and students appeared to really enjoy the negotiations.

For the third marking scheme the groups again produced the schemes from scratch and the spreadsheet was created in the same manner as before. Due to time constraints the spreadsheet and a proposed scheme were posted on Moodle and feedback on them was requested. Students commented online on the proposed marking scheme and some minor modifications to the final scheme were made.

## *2.4 Implementation of peer-assessment*
At the presentations each student was given a copy of the co-designed presentation marking scheme and asked to review a certain group (each student reviewed a single group and not their own). Unfortunately, due to time constraints the marking schemes were given to the students at the start of the presentations and thus they had minimal time to review the final scheme and prepare. The output of this process was approximately 5 to 6 reviews per group. There was a requirement that each reviewer had to ask a question to a member of the group presenting that had not answered a question previously, note their answer and the rate the quality of the answer was also added. Such an approach required each presenting student to demonstrate knowledge and facilitated the development of analytical and communication skills. The same process was followed by the lecturer with each group being marked by them. The combined peer marks, lecturer marks and two colleagues' marks formed the final mark; with the lecturers mark more heavily weighted.

### *2.5 Implementation of self-assessment*

Students completed a structured learning log upon completion of the implementation tasks (Task 1 to 4) and some of the documentation. The goal was to allow them to reflect on their own experience and on their perception of the group experience. The same learning log template was used for each task. The log can be found in Appendix A.

### *2.6 Action Research Cycles*

The action-research project was composed of a single cycle, however part of this cycle was sub-divided into two parts. The decision to use a single cycle was based upon the limited time available for the project (seven weeks). It was felt that a single cycle of a reasonable duration would lead to more reliable and authentic feedback than two or more short cycles which failed to give the students sufficient time to get used to the novel assessment structure. In addition, the sheer volume of work and the amount of survey data generated meant that by the time a change could be made it would be too rushed. The single complete assessment cycle is composed of:

- self-assessment (no changes were made during the cycle)
- peer-assessment (no changes were made during the cycle)

The co-assessment was composed of two cycles:

- Cycle 1:development of marking scheme for presentations
- Cycle 2: development of marking schemes for tasks (Task 1 to Task 4).

# 3. Findings, results and analysis

The goal of this project was to implement an *innovative assessment structure, using peer-, self-, and co-assessment,* for a *group based real-world project,* that was deemed *fair and reasonable* and provided a good *learning environment*. This section presents findings on how well this goal was satisfied with particular attention to the highlighted keywords in the above goal statement.

Several instruments were used to collect data:
- Survey carried out at the half-way point in project (provided in Appendix B)
- Survey carried out at project completion (provided in Appendix C)
- Student interviews
- Peer and mentor feedback (provided in Appendix D)

Results are provided in several parts based on the assessment items which were evaluated. Specifically, findings on each of the following are provided:
- Feedback on co-assessment
- Feedback on self-assessment
- Feedback on peer assessment
- General feedback on assessment structure
- Feedback on group work and the learning environment

## 3.1 Co-assessment feedback
*Cycle 1 Evidence:*

- Half-way survey question
    - *Did you feel the assessment scheme, built from your suggestions, was fair?* (See Appendix E).
- Class Discussion

*Cycle 2 Evidence:*

- End of project question
    - *What are your thoughts on co-assessment structure? List at least one positive and one negative.*

**Cycle 1:**

At the half-way point survey results indicated that 88% of students thought that the assessment scheme was fair with a further 8% indicating that it was somewhat fair, see Figure 2 for further details.

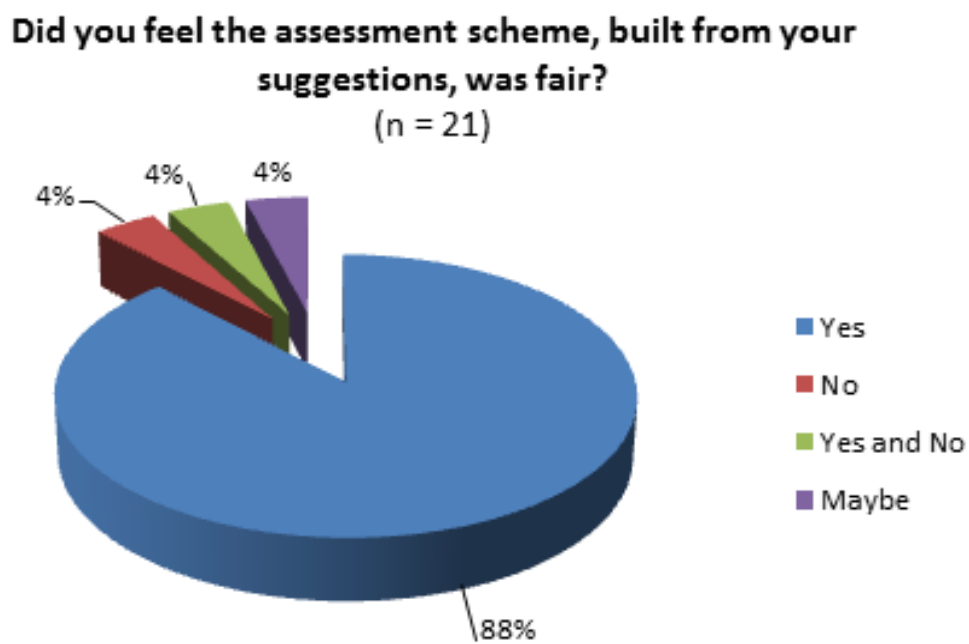

Figure 2: Findings on question 'Do you feel the assessment scheme built from you suggests, was fair?'

Sample comments given by students included (key phrases are highlighted in bold):

- 'Yeah because *it took everyone's views into consideration*'.
- 'The headings which were assessed were comprised of headings from the whole class, so yes, the *assessment scheme was fair*'.
- '…there was a *broad range of criteria* used that *judge our sites fairly*'.
- '…created a *room for the students to assess other students* based on their performances…..*some students might not want a particular group to do better than them*'.

Subsequent classroom discussion revealed that while the students liked the process and appreciated being involved they felt that they should get a chance to view the final marking scheme before using it as a reviewer. This would allow them to become familiar with the final scheme and formulate questions for the groups at presentations. No other changes were suggested. It was felt that no further change was necessary given the 96% of the students who replied found the assessment scheme to be predominantly fair and this was a substantial part of the goal achieved. As the students would not be involved in the marking of the next two co-designed schemes no change was needed to the subsequent cycle. However, in future it would be worth being conscious to return the marking scheme more promptly to the students where they are using it as reviewers.

**Cycle 2:**

Upon completion of the two other marking schemes students were asked *'What are your thoughts on co-assessment? List at least one positive and one negative*. A review of the comments indicated that no student declined to give a positive comment but several students provided no negative comments. In addition, the positive comments are subjectively far more encouraging and compelling than the negative ones. Sample comments are provided in Table 1 with a detailed list given in Appendix F – it is important to note that Table 1 and subsequent tables tend to show equal numbers of positive and negative comments to give the reader a sense of the types of comments received. However, feedback from this project was predominantly positive and the tables should be read with that understanding.

| Positives: | Negatives: |
|---|---|
| *'Good that if you identify an area as important that you can get marks for that'.* | *'Bad if people suggest silly stuff like jazz hands'.* |
| *'Student input increases interest'.* | *'Time taken from lectures'.* |
| *'Good. Positive: marking scheme is fair'.* | *'…takes up extra class time'.* |
| *'It's very good – it allows input on what we thought was important'* | *'…some people could be negative in their marks even if the group did well'.* |
| *'You see where the marks go, and why they go for each part'.* | *'May not be taken seriously'.* |
| *'I thought it was a really good idea because it was a lot fairer way of marking'.* | *'Can be biased'.* |
| *'Good to know what is being assessed'.* | *'Well, it just take time to do assessment'.* |
| *'Involves students so they pay attention'.* | *'Takes a lot of time to decide on it'.* |

Table 1: Comments on co-assessment

Finally at the end of the project students were asked if co-assessment should be maintained on the module and if more co-assessment should be introduced on the degree. Eighty percent of students felt that co-assessment should be maintained on the module but only 53% felt more should be introduced (see Figure 3).

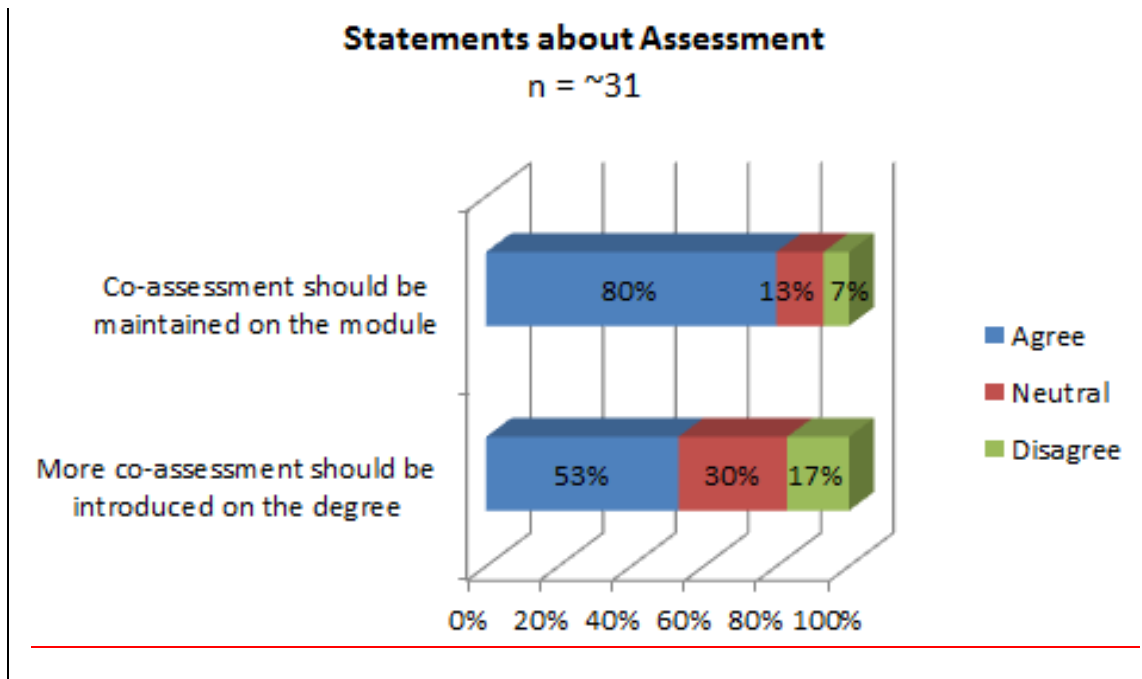

**Figure 3: Findings of statements on co-assessment**

Students believed that whilst co-assessment suited the current module it would not fit well with more traditionally taught modules that focused on individual learning.

Overall it appears that the students enjoyed and valued the co-assessment but in future better time management is needed to quickly return the schemes to students where they are to act as reviewers.

### *3.2 Self-assessment feedback*
Evidence on students' perception of the learning logs was obtained from the following sources:

- Likert-scale questions from the end of project survey:
- End of project survey question:
    - What do you think of using learning logs? List at least one positive and one negative.

The goal of the learning logs was to give students an opportunity to reflect on their role in the project and on their group work experience. Interestingly, as depicted in Figure 4 students' ratings of the value of learning logs in terms of usefulness, helping reflection and identifying strengths and weaknesses was far greater for the group (on average 68% responded positively) than for the self (on average 56% responded positively). This is interesting as it suggests that the learning logs that were designed have more value for reflecting on the group experience than on the individual experience.

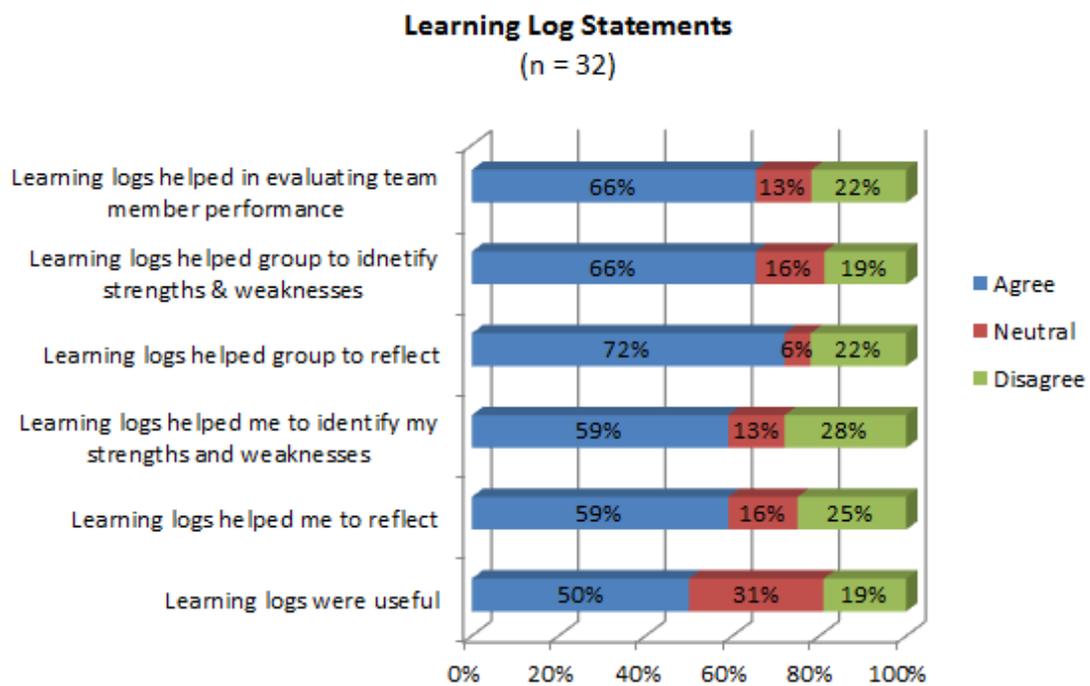

Figure 4: Findings of statements about the value of Learning Logs

To further explore the value of learning logs students were asked *'What do you think of using learning logs? List at least one positive and one negative point.'* Some sample comments are provided in Table 2 (full list in Appendix G). The principal concerns with the learning logs appears to be the length of time it takes to do them, the loss of time for items perceived to be more important, and the rigid nature of the template. These issues are reasonable and need to be dealt with for future cycles of this work. The implications are discussed in detail in the next section.

| Positives: | Negatives: |
|---|---|
| 'Helps track work'. | 'Freeform would be better…' |
| '… a good way to reflect on what you have learned through the module.' | 'I see no point to them'. |
| 'I think the learning logs helped me to assess my own progress, plus the group'. | 'It takes up time that could be spent on other tasks'. |
| 'It's *easier to notice problems* in the group'. | 'Bad thing was that *learning logs were the same for every task…*'. |

**Table 2: Comments on learning logs**

### *3.3 Peer-assessment feedback*

Evidence on peer-assessment was gathered from the following sources:

- Half-way survey question
    - *Did you enjoy being a reviewer at the presentations?*
    - *Would you prefer if only the lecturer assessed the work and classmates did not have any say in the marks?*

Shortly after the presentation students completed the half-way survey. Students were asked '*Did you enjoy being a reviewer at the presentations?*' Seventy-one percent replied 'yes', 19% replied 'maybe' or 'no preference'. The remaining 10% did not enjoy being a reviewer (refer to Figure 5). A full list of comments given by students to this question is provided in Appendix H, including:

- 'Yeah it was good to **critically review** how other teams are doing…'
- 'It was **fun**!'
- 'Yes, as it **allowed me to really engage** with their website and **forced me to really think** about it'.
- '**…it was very fair**'.

- '…it was an ***interesting experience*** and it was quite ***enjoyable. I understand better what it's like being a reviewer***'.
- 'I wasn't sure if I could ***come up with a relevant question to ask***….

The two students who replied 'no' did not give a reason for their response.

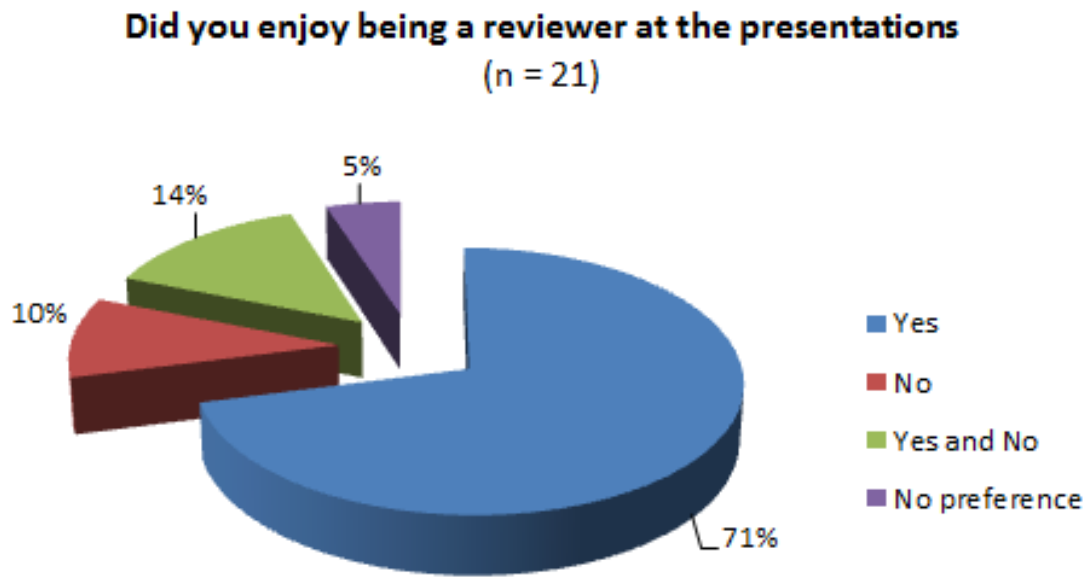

**Figure 5: Findings on question 'Did you enjoy being a reviewer at the presentations?'**

The students were also asked *'Would you prefer if only the lecturer assessed the work and classmates did not have any say in the marks?'* Results are depicted in Figure 6.

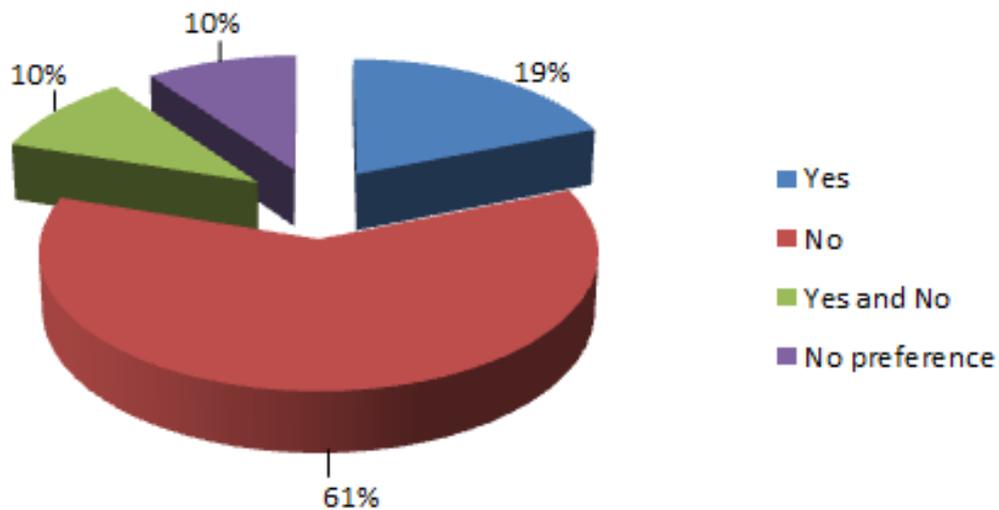

**Figure 6: Findings on question 'Would you prefer if only the lecturer assessed the work and classmates did not have any say in the marks?'**

Sixty-one percent of the students responded unequivocally negatively and 19% responded unequivocally positively with the remaining 20% expressing a mixed response or no preference. The comments provided valuable clues to the polarity of the responses, with sample comments including:

- '…classmates should **give the tutor some suggestions**, but it should be **only tutor who marks all groups**, as it is probably the **most experienced person** and also has **no interests in it (marks fair)**'.
- …everyone is going to **give a fair assessment of how other teams have performed**'.
- 'No. **Being assessed by classmates is much better as they have a similar knowledge level**…'
- 'No **I trust my peers**'.
- 'No, but I think **lecturers mark should have more weight**'.
- 'No, I like the system we used, it was **very good, very fair**'.

The complete list of comments is given in Appendix I and these findings are discussed in detail in the Action Implications section.

### *3.4 General Feedback on Assessment Structure*

Half-way through the project and at the end of the project the lecturer met with each group to evaluate how they were getting on, get feedback and discuss / resolve any issues. Some of the statements made by the groups about the assessment structure were. Sample comments pertaining to assessment included:

- *If it could be applied it should be used on other projects but not for individual work.*
- *Perfect for this module.*
- *Liked it but decide on marking scheme earlier – people need awareness that this is coming.*
- *Liked to have opinion heard but need something (template) to start with.*
- *Liked being involved in assessment, felt it was fair. Know where marks are going.*
- *Really good idea – makes you feel more involved.*
- *Showed respect to our opinion. Stayed more focused. Maybe too much of it. Adds a bit of humour – participating in it.*

To summarise, the findings indicate that students felt very positively about the co-assessment. They enjoyed and appreciated being involved in the assessment structure. Students enjoyed the peer-assessment and felt they learnt extra skills out of being a reviewer. They noted that the process helped them to engage better and remain more attentive to the presentations than if they had a less active role. Although the majority of students felt that student marks should count some concern was expressed that this could lead to bias. The suggestion that marks should count but the lecturer's should be more heavily rated was followed in this study and will be maintained going forward. The findings on the learning logs are less positive. They appear to have value for reflection but the template as it exists needs to be re-designed if the focus is on reflection of the self as opposed to reflection of the group experience.

### *3.5 Final remarks*

As integrating the new assessment structure was tightly integrated with the implementation of ***a group-based project*** and on providing a ***good learning environment*** it was important to evaluate what the students thought of both. With regard to the group-based project an indicator of whether the students were 'buying in' is given by their attendance at the group-based (lab) sessions. As can be seen in Figure 7 attendance for each group at the lab sessions was in excess of 80% for each group[1]. This suggests at least some level of involvement in the project. At the mid-point survey students were asked if there was *value in group based work.* Ninety-two percent of students recognised that there was value in it and the remaining 8% stated that it had strengths and weaknesses. A complete list of comments provided by students on this question is given in Appendix J.

---

[1] Excludes students who were moved from the group project to an individual project largely due to their lack of attendance.

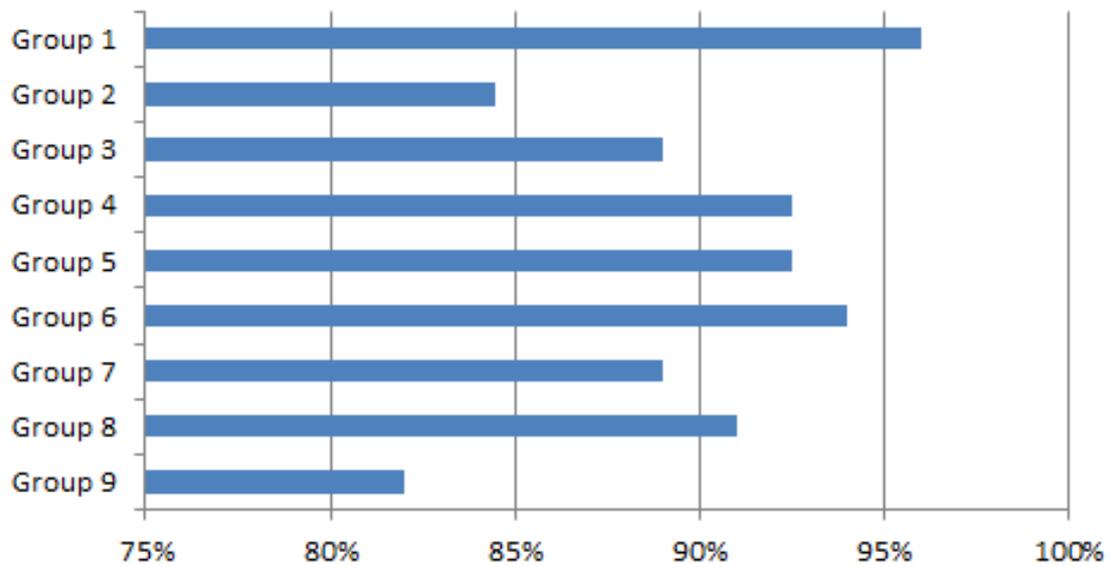

**Figure 7: Lab attendance**

At the end of the module students were asked six likert scale questions about their experience with group work. It is important to point out that the students had little or no previous experience of academic group-based projects. The statements and responses are detailed in Figure 8. Overall the findings are very positive. Of particular importance are the findings that 88% enjoyed working in a group and 78% felt they learnt more from group work.

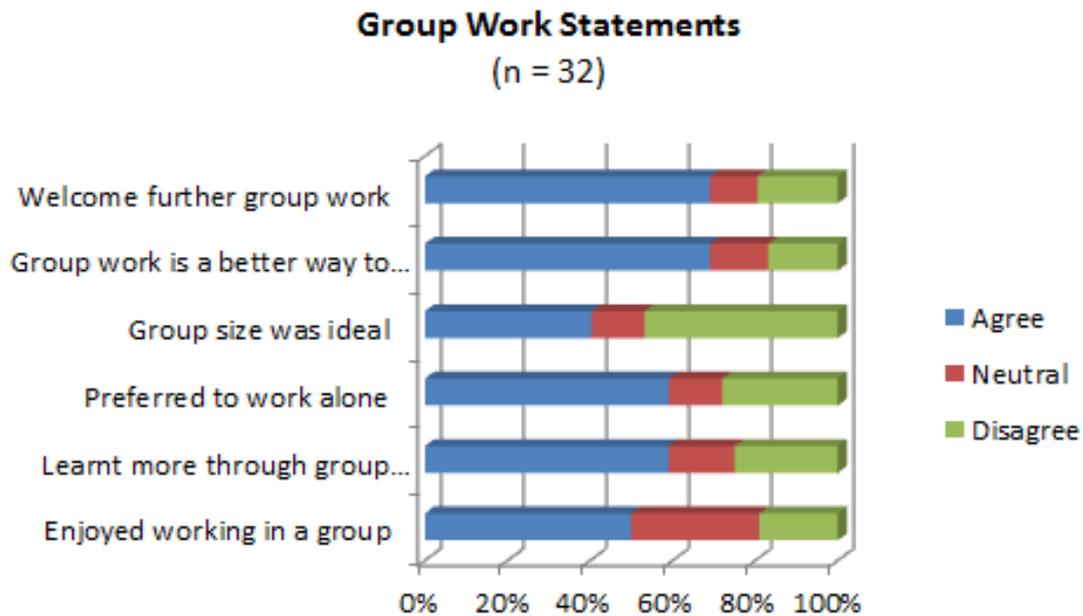

**Figure 8: Findings on statements about the value of group-work**

In addition, upon module completion students were asked *'What are your thoughts on group-based work? List at least one positive and one negative'*. Samples of their comments are provided in Table 3 (full list is provided in Appendix K).

| Positives: | Negatives: |
|---|---|
| '*You can learn new skills like leadership*'. | '*Relying on people for things and they don't deliver*'. |
| '*I enjoyed working as part of a group. I think I learned more from my team mates*, that if I did the work by myself'. | '*Felt intimidated that I knew a lot less* about code'. |
| '*More ideas* available'. | '*Timetable clashes*'. |
| '*Realistic work environment, better quality output, exposure to other ideas*'. | |
| '*We all learn more*'. | |

**Table 3: Comments on peer-assessment**

The findings suggest that the students largely enjoyed the group based work and were able to identify some of its benefits such as improving life-long learning skills, higher quality work and learning from others. Problems with communication within the group appear to be somewhat problematic with timetabling constraints making it difficult to schedule meetings during the day. The age-old issue of having to rely on other people is still a concern for students and also is the perception of not knowing enough but with 88% of students stating that they enjoyed working in a group these issues seem quite minor.

To evaluate student's perception of the learning environment students were asked a number of likert-scale questions at the end of the project (see Figure 9). As can be seen, 97% of students felt (1) there was a good learning environment, (2) the module objectives were clear and (3) the lecturer demonstrated her expertise in the area. One-hundred percent of students felt the lecturer had good control and that the material was well presented prepared while 93% felt the material was well presented.

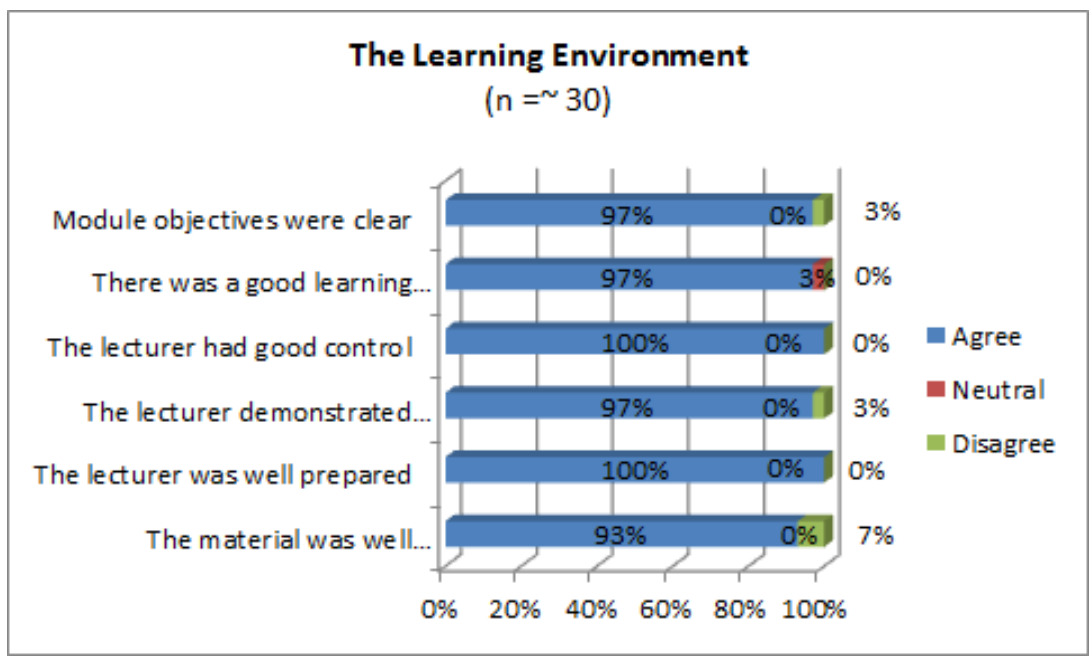

Figure 9: Findings on statements about the learning environment

Finally, at the end of the project students were asked 'Did you enjoy your experience of this module? Please explain.' Ninety-six percent of students said they did and some sample comments included:

- 'It was really helpful to work in a group as if I wasn't sure of some aspects of the course, other people in the group helped out'.
- 'Yes, group work motivated me to produce a higher standard of work'.
- 'Best module I've done so far. Enjoyable but a lot of work and work outside of lecture hours was needed'.
- 'Break from the norm. Really enjoyed a hands-on experience of project'.
- 'Yes the interaction made it stand out above my other lectures'.

The last evidence gathered on the learning environment and the group work was peer and mentor feedback. The testimonials provided are very positive and can be found in Appendix D.

# 4 Action Implications

In this section a description of the primary recommendations for colleagues wishing to engage in similar work is provided.

## *4.1 Recommendations on Co-Assessment*

The overall goal of this project was ultimately to come up with an assessment structure that the students felt was fair, reliable and promoted a good learning environment. The use of co-assessment was positively received by students and this research recommends its suitability for group-based projects. Allowing students to influence how marks are allocated helps them to feel more in involved and the vast majority of students perceived the co-designed marking schemes to be fair (88% of students thought that the assessment scheme was fair). This is important for group-based projects where student marks depends on their own and their group's performance.

However, the marking schemes take considerable time to co-develop and this reduces the amount of time available for actual curriculum content. Several students commented that although they enjoyed and valued being part of the assessment process they felt it took up precious time that could have been better spent on module material. To alleviate this, it should be explained to students at the start of the project that this exercise will allow them to develop skills such as analytical and communication skills. These skills should be incorporated into the learning outcomes so that students can appreciate their value and view them as specific curricular items.

Collating the schemes produced by each group is a large task. For this project, it took between 2 and 3 hours per collation. For busy academics finding the time to invest in this work is difficult. An interesting solution might be to get each group to work together initially to produce a marking scheme and upon completion work with 2 other groups to produce a meta-collated scheme. For a similar sized project this would mean the teacher would then need to collate only 3 schemes. Additionally, this would allow groups to work with other groups and further develop their communication and negotiation skills.

Summary of recommendations:

- Students perceive co-assessment to be enjoyable and fair on group-based projects. Based on this, the use of co-assessment is strongly recommended on similar projects.
- Students need to appreciate that co-assessment helps them to develop skills that are an important part of their learning. This should be clearly explained to students and incorporated into the learning outcomes.
- To reduce workload and further enhance student learning, groups could work with other groups to generate one marking scheme per every three groups, thereby reducing the teacher's workload and providing them with an opportunity to work with different students.

### *4.2 Recommendations on Self-Assessment*

The goal of the learning logs was to give students an opportunity to reflect on their role in the project and on their group work experience. The structure of the current template appears to help groups to reflect better on their experience than to help individual students to reflect on their personal experience. Students noted a number of concerns with using learning logs including the length of time it takes to do them, the loss of time for items perceived to be more important, and the rigid nature of the template. As a result, going forward this project will move away from the rigid template structure for self-learning logs but maintain it for group learning logs. Instead, individuals should keep a free-form reflective diary. The diary should be written up in their own time and checked weekly by mentors. Such a process should provide students with more flexibility but also encourage them to be timely in their entries. This is also important from the teacher perspective as the current system resulted in nearly 300 logs that had to be reviewed by the teacher. Such a system is not sustainable but having a mentor review twenty short journal entries on a weekly basis, with the teacher randomly sampling their reviews would be achievable. Furthermore, this will help mentors to develop critical thinking and communication skills and make them more aware of individual issues that need to be dealt with.

It is important that students receive some direction in how to reflect at the start of the year. This project found that students at this stage in their learning need a template / guide as a starting point.

Summary of recommendations

- Maintain existing learning log structure to help promote group reflection.
- Incorporate a free-form reflective diary, with guidance, for individual reflection. The diary should be checked regularly and frequent feedback should be provided.

### *4.3 Recommendations on Peer-assessment*

Peer-assessment worked well in this project with 71% of students indicating that they liked being a reviewer. Student comments included phrases like *'it was good to critically review'*, *'it was fun!'*, *'it allowed me to really engage'*, and *'forced me to really think'*. This is in line with previous findings on the value of peer-assessment and its use on other modules is strongly promoted here.

Peer marks should be included in the assessment structure but should receive a lower rating that that of the lecturer and other experienced staff. As outlined in Section 3.3, 20% of students who expressed a mixed response would be largely be satisfied by this, which would result in over 80% agreement that peer marks should be incorporated.

However, incorporating peer marks involves a large amount of extra work for the teacher. In this project there were approximately 45 sets of peer marks to be taken into account (approximately 5 peer marking schemes per group). Again, for a busy academic this is considerable additional work and its sustainability is questionable. It is difficult to know how to reduce this workload aside from incorporating only a small random number of marks (say 2 per group). A review of the schemes from this work suggests that the marks awarded by peers to the same group were highly correlated which suggests this may be a reasonable solution.

Summary of recommendations:

- Peer assessment is valuable for learning and should be used on other modules.
- Peer marks should be included in final marks but with a lower weighting.
- Where a large amount of peer marks are generated using a small random sample of these marks may be appropriate.

### *4.4 General Recommendations*

- Students really enjoyed working in teams and felt they learnt more from group work. They appreciated the value of learning from their peers and felt motivated to produce better work. Furthermore, they were able to identify that it helped them to develop lifelong learning skills such as

- critical thinking. It is a recommendation of this research that group work should be incorporated more on the syllabus.
- Students appreciate different learning environments. They valued the break this module gave them from more traditional teaching structures and in particular valued the amount of interaction it involved. As teachers, we should feel more confident that taking educated risks in the classroom can result in a better and more appreciated learning environment for our students
- Finally, on a personal note, although any change made in the classroom usually requires more work for the teacher the sense of satisfaction felt upon completion is worth it.

# 5 Conclusion and Final Remarks

Students perceived the learning environment to be very good, enjoyed the group-work, the co-assessment, the peer-assessment and to a lesser degree the use of learning logs. The feedback from the students was for the most part positive and the minor negative issues can be dealt with now the foundations have been laid for this new approach.

The quality of the websites built by the students were excellent. Numerous colleagues commented on the high level of student engagement and the technical level achieved by a largely second-year student group.

That said, the work load to implement this process is substantial. This project generated approximately 300 learning logs that had to be read and marked; not to mention a large amount of time spent on collating; and designing the marking schemes and the subsequent integration of the students' presentations marks with colleague marks and my own. However, a free-form diary that mentors will review on a weekly basis and is randomly reviewed by the lecturer would significantly reduce the work load. As for the collation and integration of marks, automation, as least in part of this process would again alleviate some of the work.

Tucker, R., Fermelis, J. and Palmer, S. (2007) Online self-and-peer-assessment for teamwork in architecture and business communications ANZASCA 2007: Proceedings of 41st annual conference. Towards solutions for a liveable future: progress, practice, performance, people. 264 – 271

## Appendix A: Learning Log Template

Learning Log Template (ALL SECTIONS MUST BE COMPLETED)

- **Title**
  - The name of the task that the entry relates to
  - Your name and student number

|   |
|---|
|   |

- **Context**
  - What the task involved

|   |
|---|
|   |

- **Your role**
  - What you did on the task
  - How long you spent working on it.

|   |
|---|
|   |

- **Your experience**
  - What you learned from this experience, specifically outline
    - the routine - what you expected to discover
    - the unexpected – what you learnt that surprised you

|   |
|---|
|   |

- **Your views on the group**
  - How well did the group collaborate on the task

|   |
|---|
|   |

- **Group Work Positives**
  - What advantages you have identified from working in your group

|   |
|---|
|   |

- **Group Work Difficulties**
  - What difficulties you see emerging and how you propose to deal with them

|   |
|---|
|   |

**Marks Distribution (your perception of a fair distribution of the marks).**

Outline how you believe the marks for the work completed up to the interim presentation should be apportioned. You cannot assign the same mark to more than three people. See the example below.

Example:
Assume the team is composed of six members: Tom, Dick, Harry, Ann, Jane and Mary (yourself). Assume the group receives an overall mark of 70, which is a total of 420 marks for the six team members. You need to apportion the marks with the following restrictions
    (1) Not more than three team members can get the same number of marks
    (2) No team member can get more than 100 marks.

In your opinion, Tom and Dick did a huge amount of work, Harry and Ann did a lot of the work but Jane was absent and did very little. You feel you also worked hard, as hard as Harry and Ann. The following grid could represent how the marks should be apportioned. By re-distributing the 360 marks Tom and Dick get the most reward, followed by Harry, Ann and you, with Jane receiving a fairer mark based on her contribution. N.B. This is just a sample scenario, you need to think about how your group performed and assign what you believe to be a fair mark to each member.

**Assume your group has also received 70 marks.**

| Name | Share of marks |
| --- | --- |
| Tom | 85 |
| Dick | 85 |
| Harry | 70 |
| Ann | 70 |
| Mary | 70 |
| Jane | 40 |
| **Total** | **420** |

Complete your group's grid here:

| Name | Share of marks |
| --- | --- |
|  |  |
|  |  |
|  |  |
|  |  |
|  |  |
|  |  |
|  |  |
| **Total** |  |

## *Appendix B: Midpoint Feedback Sheets for Module*

CS230 Feedback (Upon completion of Tasks 1 and Tasks2)

1. Are you enjoying your experience of this module? Please explain.

2. Would you prefer if the older approach was still used? Please explain.

3. Do you think there is value in group-based work. Please explain.

4. What could be done to improve this module?

5. Did you enjoy being part of the assessment structure for the presentations? Please explain.

6. Did you feel the assessment scheme, built from your suggestions, was fair?

7. Did you enjoy being a reviewer at the presentations?

8. Would you prefer if only the lecturer assessed the work and classmates did not have any say in the marks?

9. Please provide any other comments that would help me to make improvements.

## *Appendix C: End of Module Feedback Sheet*
COMPUTER SCIENCE STUDENT FEEDBACK FORM
Course CS230
Lecturer : Dr. Susan Bergin

Your responses are anonymous. **Do not write your name or student ID number on this form**.

Enter today's date here (day/month/year) : ______/______/______.

**Part I : Give your feedback for statements 1-25 on a scale of 0 to 4 as follows :**
[4] = Strongly Agree
[3] = Mildly Agree
[2] = Neither Agree or Disagree - (or) - Not Applicable
[1] = Mildly Disagree
[0] = Strongly Disagree

|  | Statement | Score |
|---|---|---|
| The Module | 1. The module objectives were clear | |
| | 2. The module has significantly increased my understanding of web development | |
| | 3. The course has significantly increased my understanding of CS in general | |
| | 4. It was evident how topics covered on the module related to one another | |
| | 5. There was sufficient time to cover all the module material | |
| The Group Work | 6. I enjoyed working in a group | |
| | 7. I learnt more by working in a group | |
| | 8. I would have preferred to work alone | |
| | 9. The group size was ideal | |
| | 10. Group work is a better way to learn where appropriate | |
| | 11. I would welcome further group work where appropriate | |
| The Learning Logs | 12. The learning logs were useful | |
| | 13. The learning logs helped me to reflect on my experience | |
| | 14. The learning logs helped me to identify my strengths and weaknesses | |
| | 15. The learning logs helped my group to reflect on my groups experience | |
| | 16. The learning logs helped the group to identify strengths and weaknesses | |
| | 17. The learning logs were useful in helping members of the group to identify how the team members including themselves were performing | |
| | 18. The learning logs were useful for other reasons | |
| | 19. The learning logs should count for marks | |
| Moodle | 20. I enjoyed using the forums on Moodle | |
| | 21. I enjoyed using the chatroom on Moodle | |
| | 22. My group used Moodle to communicate outside the lab times | |
| | 23. My group communicated by email or in person outside the lab times | |
| | 24. Communication between group members was a problem | |
| Assessment | 25. I enjoyed being part of the assessment process (co-assessment) | |
| | 26. It is more respectful to ask students their opinion on how the marks should be broken up for group based software projects | |
| | 27. The marking scheme build from our suggestions were fair | |
| | 28. Co-assessment should be maintained on this module | |
| | 29. More co-assessment should be used on the degree program | |
| The Lectures | 30. There was a good learning environment | |
| | 31. It was easy to hear the lecturer | |

|  | 32. It was easy to read what was presented in the lectures | |
| The | 33. I found the tutorials useful | |
| Tutorials | 34. I felt there were enough tutorials | |
| The Lecturer | 35. The lecturer had good control of the class | |
|  | 36. The lecturer demonstrated his/her expertise in the area | |
|  | 37. The lecturer seemed well prepared for each lecture | |
|  | 38. The course material was well presented in the lectures | |
|  | 39. The lecturer answered questions satisfactorily | |
|  | 40. The lecturer was available for questions outside class time | |
|  | 41. The lecturer increased my interest in the subject | |

1. Did you enjoy you r experience of this module? Please explain.

2. What are your thoughts on group-based work? List at least one positive and one negative.

3. What do you think of using learning logs? List at least one positive and one negative.

4. What are your thoughts on co-assessment structure? List at least one positive and one negative.

5. What parts of this module did you find easiest to understand?

6. What parts of this module did you find hardest to understand?

7. What aspect of this lecturer's teaching did you appreciate the most?

## *Appendix D:* Peer and Mentor feedback

***Dr. James Power (Lecturer – CS Department):***

'In the current academic year I have acted as mentor for one of the demonstrators on Susan's CS230 module in second year, and this has given me the opportunity to observe both demonstrators and students in the labs for this module. A number of examples of best practice stand out here. First is the high volume of feedback supplied to students at all stages during the module. Second is the emphasis on well-integrated teams, in a class of students from diverse educational backgrounds. Third is the substantial technical agenda: the students designed, developed and implemented a multimedia web site during this module. I observed the students in their final lab and for their project presentations and was particularly impressed with their positive attitude, confidence and sense of pride in their success during this module.'

***Mr. Patrick Marshall (Technician, CS Department – responsible for Departmental Web System):***

'Judging from the end result of this module, I feel that the students have gained invaluable experience in web development. The final product that the students delivered seemed well beyond what I would have expected from a second year module. The presentations were delivered in confidence as they were developed along with their projects throughout the module. Susan has guided the students through a professional web development cycle, which allowed tasks to be divided up amongst students from various subject areas. This maps directly to industrial practices where multifaceted disciplines join together to create an end product. So the overall cycle of team work, task management, final presentation and deliverable, will benefit the students greatly going forward'.

*Dr. Aidan Mooney, Lecturer – Department of Computer Science, NUIM:*

'I got the feeling that the students thoroughly enjoyed working on this and liked the challenge of working in competition with the other teams. There was a general good feeling in the groups between the members and it was well structured so that there was a nice mix of disciplines in the teams. The majority of presentations were excellent and the web pages generated are to a very high standard'.

*Amy Fitzgerald CS230 group project mentor:*

'I enjoyed demonstrating this module as it was well-organized, there was a positive atmosphere in the lab each week and learning outcomes were clear and constantly assessed. I have listed the positive and negative aspects of the labs for this module below.

**Positive**

- **Enthusiasm:** Students were enthusiastic about the labs and the module as a whole because their feedback and input was taken seriously and helped to shape the course. E.g.: marking schemes
- **Team Building Skills:** Team members were co-operative with each other and hard working. I think this is because of the mid-module review where people could choose to work individually, the conscious choice caused students to put in more effort and pull their own weight
- **Practical Computer Science Skills:** Students were learning a practical profitable skill so it was easy to motivate them to learn

- **Demonstrator Feedback:** As a demonstrator I felt that my feedback was taken into account when planning the labs
- **Attendance:** As the students were working in teams this encouraged high attendance
- **Personal Skills Development:** Not only did the students learn web development they also learned team management and presentation skills

**Negative**

- **Individual Projects:** Students who opted to work on individual projects had very poor attendance in the lab and so did not receive as much guidance as the students who worked as part of a team
- **Complacency:** A relaxed atmosphere is a great thing but sometimes I felt that they did not take some of the tasks seriously, such as the second marking scheme'

*Danny Fallon - CS230 group project mentor:*

'Demonstrating this module was great. The labs were well-organised, with clear goals and lab assignments each week on moodle. Below is my list of positive & negative points:

**Positive**:

- **Team Building Skills.** The team aspect of the labs helped in many ways, including preparation for the Team Project module in 3$^{rd}$ year CSSE. Throughout the course of the module the various teams discovered just what it's like to be part of a team, including taking other team members input into account, realising they can't do it all themselves. The opportunity to rate each other really improved how the

teams got on together and made sure (in most cases) that the workload was distributed evenly

- **Real-World Product.** At the end of the day the teams produced a working website. Perhaps not the most sophisticated website ever, but the module definitely gave them the chance to feel out if they would like to explore web development as a potential specialisation which was really good.

- **Appealed to all disciplines.** CSSE students were not the only participants in the module. Task 4 allowed the students who weren't the most comfortable when digging around code to still use their other skills (e.g. the Media students did a lot of the video editing in my groups).

- **Presentations & Course Feedback.** As a group the students were able to come up with marking schemes for tasks and rate them. This is similar to putting a framework together out in the real IT/Business world to decide on what option to go with – a real skill. Then there were the presentations, from the requirements analysis to the actual product, all setting them up for numerous things including final year and beyond.

**Negative:**

- **Enthusiasm & Complaceny:** Maybe it was the subject matter for the website, maybe it was a lack of web development drive in general, but one thing is sure – the students were pretty complacent in the labs. On numerous occasions I was getting marking schemes handed to me scribbled on scraps of paper. Girlfriends coming into labs, students spending half of their time on bebo/facebook/games/other websites instead of contributing to the project as a team etc. etc. As much as they were reminded what they were in the lab for, we're not babysitters so I was pretty surprised to see so much time wasted doing anything other than CS230 work. This made the labs and team progress hard to judge, because a lot of the work was done outside of our 2 hours.'

## *Appendix E: Assessment Scheme Feedback*
Did you feel the assessment scheme, built from your suggestions, was fair?
**Yes:  20    No: 1      Both 1          Maybe 1**

**YES**

'Yeah because it took everyones views into consideration'.

'Yeah because it included everything'.

'The headings which ere assessed were comprised of headings from the whole class, so yes, the assessment scheme was fair'.

'Yes I feel that was fair'.

'Yes, choosing the best and most common questions from each group makes for a fair scheme'.

'Yes, I did'.

'Totally, I thought it was fine'.

'Yes, I agree it was fair!'

'Yes! It was also well selected by our lecturer things that were good were included and things that perhaps were less relevant e.g. jazz hands were excluded. Although I think jazz hands would have been a very cool one to assess on'.

'Yes I do as it included everything I thought we needed to be assessed on'.

'Yes'.

'Yes'.

'Yes'

'Yes, it was fair and balanced'.

'Yes'.

'Yes, I think it was fair'.

'Yes, I though it marked relevant section in order of importance'.

'This is democracy'.

'Yes. There was a broad range of criteria used that judge our sites fairly'.

'Yes it covered all necessary areas with proportionate marks.

**NO**

No, probably not.

**BOTH**

'YES! It created a room for the students to assess other student based on their performances. No! Some students might not want a particular group to do better than them, and for this reason mark them down'.

**MAYBE**

'Perhaps, I'm jus afraid of any bias coming from people'

## Appendix F: Co-Assessment Structure Feedback

EOY: What are your thoughts on co-assessment structure? List at least one positive and one negative.

**+:**

'Good that if you identify an area as important that you can get marks for that'.

'One positive aspect of co-assessment is that we can (kind of) control what marks we can achieve'.

'Student input increases interest'.

'Good. Positive: marking scheme is fair'.

'It's very good – it allows input on what we thought was important'

'Sweet'.

'You see where the marks go, and why they go for each part'.

'I like it, as it allows you to be marked on what you did well'.

'I liked the co-assessment structure. I think with this structure everyone is forced to contribute and the marking scheme was fair'.

'It allowed the marking scheme to be based on what the creators of the site deemed important'.

'Every website is unique so should be assessed as such'.

'I thought it was a really good idea because it was a lot fairer way of marking'.

'Good idea'.

'Good to know what is being assessed'.

'We are marked on what we feel we should be marked on'.

'Involves students so they pay attention'.

'This was good. A positive: it allows students views to be heard'.

'Focuses you on what's important, learn what others think are important'.

'It was good we got to provide input'.

**-:**

'Bad if people suggest silly stuff like jazz hands'.

'Time taken from lectures'.

'Again, takes up extra class time'.

'One negative aspect is that some people could be negative in their marks even if the group did well'.

'Some bickering can ensue'.

'May not be taken seriously'.

'Means things like jazz hands have a hope to get in'.

'Only negative if someone doesn't do the work, they still get the same mark'.

'Frivolous marks'.

'Can be biased'.

'No negative'.

'No negative remark'.

'Marks could be styled to suit others'.

'Well, it just take time to do assessment'.

'Some students ideas may not have been used'.

'Takes a lot of time to decide on it'.

'Could lead to subjectivity'.

## *Appendix G: Learning Logs Feedback*

EOY: What do you think of using learning logs? List at least one positive and one negative.

**+:**

'Working out if someone is slacking'.

'Its useful for me'.

'Helps track work'.

'It's alright. Positive: helps identify any problems'.

'Learning logs are a good way to reflect on what you have learned throught the module.'

'You realise how much work you put in'.

'Reflecting on what you've learned'.

'It helps to reflect on what was achieved'.

'That was good because we can reflect on our experience'.

'I think the learning logs helped me to assess my own progress, plus the group'.

'Let you reflect on your strengths & weaknesses in the group'.

'Good at looking back on what we did'.

'Useful'.

'Focuses you for next task'.

'Chance to reflect'.

'Good to reflect'.

'It was good to use them. A positive: it allows us to recap on what we've learnt'.

'It's easier to notice problems in the group'.

'Good for reflection'.

**-:**

'Freeform would be better. Sometimes there is nothing to say'.

'I see no point to them'.

'It takes up time that could be spent on other tasks'.

'Bad thing was that learning logs were the same for every task, and in my opinion it shoudn't'.

'I don't like it. No positives, just a waste of time'.

'Shouldn't (spend) too much time on them'.

'Bad idea – they do not reflect job done'.

'Easy to forgot, hard to catch up on'.

'No negative'.

'They were tedious at times'.

'They were time consuming'.

'I don't think they are any negatives'.

'No positives, all negative'.

'It takes time'.

'May be more productive to spend less time on this and more on other aspects (where we still have a lot to learn).

'Didn't like it – seemed like a waste of time'.

'The temptation to lie'.

'Sometimes it was hard to remember exactly how much time we spent on any thing'

'Sometimes it is hard to fill out parts'.

'If there really was a problem someone could bring it straight to the lecturer'.

'It wasn't really work the effort sometimes'.

## *Appendix H: Student Feedback on being a Reviewer*

Did you enjoy being a reviewer at the presentations?

**Yes: 15  No: 2  MAYBE: 3  No Preference: 1**

**YES**

'Yeah it was good to critically review how other teams are doing better than us or worse than us'.

'It was fun!'

'Yes. We can ask questions about things we feel are unclear and see how well the group handles questions, especially unexpected ones'.

'Yes, as it allowed me to really engage with their website and forced me to really think about it'.

'Yes that was nice'.

'I think it is good to have some opinion from many people rather than just one person (tutor) but I did not like review some groups myself'.

'Yes. I thought it was a good idea the way it was organised – since it prevented a group from getting to friends to ask these questions – it was very fair'.

'Again I didn't mind it, definitely was interesting to be able to discuss peoples site with them, listen to what they had to say'.

'Yes! It was an interesting experience and it was quite enjoyable. I understand better what it's like being a reviewer'.

.'Yes as again we know what prospective students are looking for as were prospective students 2 years ago'.

'Yes'.

'Yes'.

'Yes it was interesting'

'Yes'.

'Yes, I believe I had a fair view of how to mark other groups'.

**NO**

'NO'.

'No, not at all'.

**MAYBE**

'I didn't mind, it feels harsh doing so for some people'.

'I wasn't sure if I could come up with a relevant question to ask, but I think I gave appropriate marks to the team I was reviewing'.

'Maybe'.

**NO PREFERENCE**

'I did not enjoy nor did I not enjoy being a reviewer'

## Appendix I: Student Feedback on Lecturer Role in project

Would you prefer if only the lecturer assessed the work and classmates did not have any say in the marks?

Yes: 4     No: 13     No Preference: 2     YES and No: 2

**YES**

'Yes'.

'Somehow yes'.

'I don't really like classmates. I think that would be better if lecturer be more involved'.

'I think classmates should give the tutor some suggestions, but it should be only tutor who marks all groups, as it is probably the most experienced person and also has no interests in it (marks fair)'.

**NO**

'No because I would feel that everyone is going to give a fair assessment of how other teams have performed'.

'No. Being assessed by classmates is much better as they have a similar knowledge level about the subject and so rather than guessing or assuming you know or understand some thing they will know'.

'In real-life, a website's success depends on what many think of it – not just one person. Allowing the class to score the presentation and website gives a more realistic score'.

'No I trust my peers'.

'I guess not, as the lecturer is subject to opinion and must have more to balance it out'.

'I think classmates would probably have a similar opinion on sites as a whole, so it wouldn't be a bad thing if their say had an impact on grades'.

'No'.

'No'.

'No the current way seems to be the best way to assess the module'.

'No'.

'No, but I think lecturers mark should have more weight'.

'No. It's good to know how our projects were judged and being able to decide on some of the marking criteria'.

'No, I like the system we used, it was very good, very fair'.

**NO PREFERENCE**

'It doesn't make difference for more'

'It is of no matter to me because I don't care assessment. I suppose it makes things more entertaining, interactive, if students assess each other, at least for those who care.

**YES and No**

'Yes and No! I did realise that we were in direct competition against all other groups so I thought it wasn't a good idea to assess ourselves. On the other hand it was a good idea because it got us a feel of what lecturers would look for and it made competition tougher'.

'Yes + no. But as its clear + transparent, then I have no problem'.

## *Appendix J: Feedback on Group-Based Work*

Do you think there is value in group-based work?

Yes   24          No          Not sure    Yes / No 2

**YES**

'Yes of course, each person can bring their own ideas and experience to the table'.

'Definitely, different people with different strengths, within the same groups makes for much easier problem solving when compared to someone tackling a problem on their own'.

'Yes. In the real world one generally works in groups a lot on projects so the group-based work gives us an idea of how well we would work in a group for a real world job'.

'Yes, as in real world situation, projects are done in teams'.

'Yes, there is less pressure on you and it is more enjoyable interacting with other team members, than doing it by yourself'

'It makes the workload manageable and easier to understand'.

'Yes because I knew nothing about coding but working with other students has helped a lot'.

'There is great value in group value in group-based work! As the saying goes, united we stand, divided we fall. Sharing of knowledge was so helpful'.

'Absolutely. In a real-life working environment there are times when group-work is essential'.

'For sure there is, but it also depends from place we are going to work in future'.

'Yeah, since it will allow me to get used to working in teams. This is more similar to any industry'.

'Yes I do: distribution of workload. However personalities on groups can be harmful. Can create friendship and foes.

'Yes but it depends on the groups of course'.

'Of course there is! Not only does it prepare us for real-world where software engineers always work in groups, but it also helps build our character and it helps us understand the aspects of working with different type of people'.

'Yes it helps you to learn to work with different personalities'.

'Yes as there are people from different study backgrounds in each group which gives a lot of different opinions and ideas for the website'.

'Yes. It makes the workload easier and it makes it more fun'.

'I think there is value in group-based work as most modules don't give an opportunity to learn to work in a group'.

'I think it's valuable as we haven't had much experience with it yet'.

'Yes, we can share some experience between each other and help each other'.

'Yes, it means that the workload is evenly distributed'.

'Yes. A lot of new skills can be learned from working with other people from different educational backgrounds'.

'Yes, learning how to work in a team is a valuable asset to have. It will help later in life'.

'Yes, but I don't like it. In the real world most software engineering projects will be group based.

**YES / NO**

'In my opinion this kind of work has both advantages and disadvantages. You have less responsibilities and work to do and you can rely on your group members and learn from each other, but there might be some conflicts in the group which can be a difficulty to finish project on time'.

'It allows (when groups are well organised) to do far more than a single person. In the other hand students are more likely to focus on only one task / technology and they might lack some knowledge at the end of the module'

## *Appendix K: End of Year Feedback on Group-Based Work*

EOY: What are your thoughts on group-based work? List at least one positive and one negative.

**+ :**

'When all members can communicate and are aware of what is going on it works really well'.

'More strengths in different areas'.

It's good. Positive: work is divided evenly'.

'You can learn new skills like leadership'.

'It's very beneficial –everyone has input and gets more work done'.

'That's easier to divide the work between the people. Less to do'.

'One positive aspect is that we all helped each other and the workload was shared'.

'Working in a group means that a person can focus on their strengths'.

'I thought the group based work was good. I enjoyed working as part of a group'.

'Not too much work'.

'I enjoyed working as part of a group. I think I learned more from my team mates, that if I did the work by myself'.

'It's a good idea, each person brings something to the group'.

'Divides work load'.

'Good'.

'Realistic work environment, better quality output, exposure to other ideas'.

'Easy to learn'.

'More time'.

'We all learn more'.

'Can designate tasks'

'It's easier to notice problems in the group'.

'The group were very helpful so I learn a lot from them'.

'Work being spread out made it possible to finish a website of higher quality'.

'More ideas available'.

'This does mean that improving your weaker skills is harder though'.

'I thought the group based work was a great idea. A positive is we learnt more'.

'On the one hand splitting up the work load made it easier…'

-:

'When students don't show up or can't do the work asked of them it can bring down the group'.

'Relying on people for things and they don't deliver'.

'Timetable clashes'.

'People can let you down'.

'No negative'.

'Have to rely on others'

'It's hard, we have different experiences and skills'.

'A negative aspect of working in a group is that there could potentially be a dominant person in the group and some opinions in the group might not be heard'.

'Communications can be hard at times'.

'No real negative'.

'One problem was trying to find a meeting that suited everybody'.

'Each person brings something to the group, but it does slightly depend on the luck of the draw'.

'The only negative was some members didn't participate fully, even though they got the work done'.

'Artistic differences'.

'Sometimes breakdown in communication can happen due to lack of clarity of roles'.

'Hard to agree'.

'Felt intimidated that I knew a lot less about code'.

'Can be very difficult to meet up outside of lecture hours'.

'When there is a communication problem'.

'Possible to let rest of group do the work'.

'Conflicts are possible'.

'... but the high variance in skill level prevented even workload'.

'A negative it is difficult to find a time everyone could meet up after class time'.